\documentclass[prx,superscriptaddress,reprint,showpacs,longbibliography]{revtex4-2}[prx]
\usepackage{color}
\usepackage{graphicx,textcomp,amssymb,amsmath,dcolumn,hyperref}
\usepackage{siunitx}
\usepackage{comment}
\usepackage{physics}

\newcommand {\bkt} [1] {\langle #1 \rangle}

\usepackage{romannum}

\begin{document}

\title{Single-shot readout in graphene quantum dots}
\author{Lisa Maria G\"achter}
\email{lisag@phys.ethz.ch}
\author{Rebekka Garreis}
\author{Chuyao Tong}
\author{Max Josef Ruckriegel}
\author{Benedikt Kratochwil}
\author{Folkert Kornelis de Vries}
\author{Annika Kurzmann}
\affiliation{Solid State Physics Laboratory, ETH Zurich, CH-8093 Zurich, Switzerland}

\author{Kenji Watanabe}
\author{Takashi Taniguchi}
\affiliation{National Institute for Material Science, 1-1 Namiki, Tsukuba 305-0044, Japan}

\author{Thomas Ihn}
\author{Klaus Ensslin}
\author{Wister Wei Huang}
\affiliation{Solid State Physics Laboratory, ETH Zurich, CH-8093 Zurich, Switzerland}

\date{\today}

\begin{abstract}
 
Electrostatically defined quantum dots in bilayer graphene offer a promising platform for spin qubits with presumably long coherence times due to low spin--orbit coupling and low nuclear spin density. We demonstrate two different experimental approaches to measure the decay times of excited states. The first is based on direct current measurements through the quantum device. Pulse sequences are applied to control the occupation of ground and excited states. We observe a lower bound for the excited state decay on the order of hundred microseconds. The second approach employs a capacitively coupled charge sensor to study the time dynamics of the excited state using the Elzerman technique. 
We find that the relaxation time of the excited state is of the order of milliseconds.
We perform single-shot readout of our two-level system with a visibility of $87.1\%$, which is an important step for developing a quantum information processor in graphene.



\end{abstract}

\maketitle

\section{Introduction}
Spin qubits in semiconductors~\cite{Loss1998,stano2021review} have the advantage that the operation and fabrication of gate electrodes are similar to classical transistors. High-quality qubits have been demonstrated on traditional bulk MOSFETs~\cite{Veldhorst2014,Yang2018,Veldhorst2015} as well as on \Romannum{3}-\Romannum{5}~\cite{Nakajima2020,Cerfontaine2014,Nichol2017}, silicon- ~\cite{xue2021computing,Zajac2018,Yoneda2018,mills2021twoqubit} and germanium-~\cite{Hendrickx2021} based heterostructures. Furthermore, semi-industrial structures compatible with industrial Si-technologies, such as fully-depleted silicon-on-insulator (FD-SOI) transistors~\cite{Maurand2016} and fin field-effect transistors (FinFETs)~\cite{camenzind2021spin}, have been investigated.

Graphene offers several advantages as a host material for spin qubits, namely naturally low nuclear spin concentrations and weak spin--orbit interactions, similar to Si. In addition, the 2D nature of graphene allows for much smaller and possibly more strongly coupled quantum devices~\cite{Liu2021}. Furthermore, bilayer graphene quantum dots (QDs) offer the flexibility of bipolar operation~\cite{Tong2021}. Compared to the mature Si-based technology, the development of quantum devices in graphene is in its infancy.
Recent advances in the controllability of individual states in single QDs~\cite{Tong2021,Garreis2021,moller2021probing,Kurzmann2021} and double QDs~\cite{tong2021pauli,Banszerus2021Spinvalley}, as well as the implementation of charge detection~\cite{Kurzmann2019Detector}, enable the realization of spin qubits based on electrostatically defined QDs in bilayer graphene.
Major milestones such as qubit manipulation and detection have yet to be achieved to unlock the qubit potential of graphene.


Single-shot readout is an essential first step towards building a universal quantum computer and implementing quantum algorithms and quantum error detection. 
In order to reach the single-shot readout limit, it is critical for the excited state relaxation time to be longer than the measurement time to resolve a single charging event. 
We first investigate the relaxation time by measuring the current flowing through a QD~\cite{Fujisawa2001,Hanson2003}, which allows us to extract a lower bound only, similar to previous experiments~\cite{banszerus2021spin}.
In order to study the time dynamics of the excited state beyond the microsecond regime, we add a charge detector to the device design and perform time-resolved measurements of the tunneling events in the QD. This allows us to perform single-shot readout of the two-level system using the Elzerman technique~\cite{Elzerman2004}. We estimate the relaxation time to be in the millisecond regime which is a few orders of magnitude longer than typical spin-qubit operation times\cite{Koppens2006, Hendrickx2020}.



\section{Pulsed-gate spectroscopy}

\begin{figure}
	\includegraphics{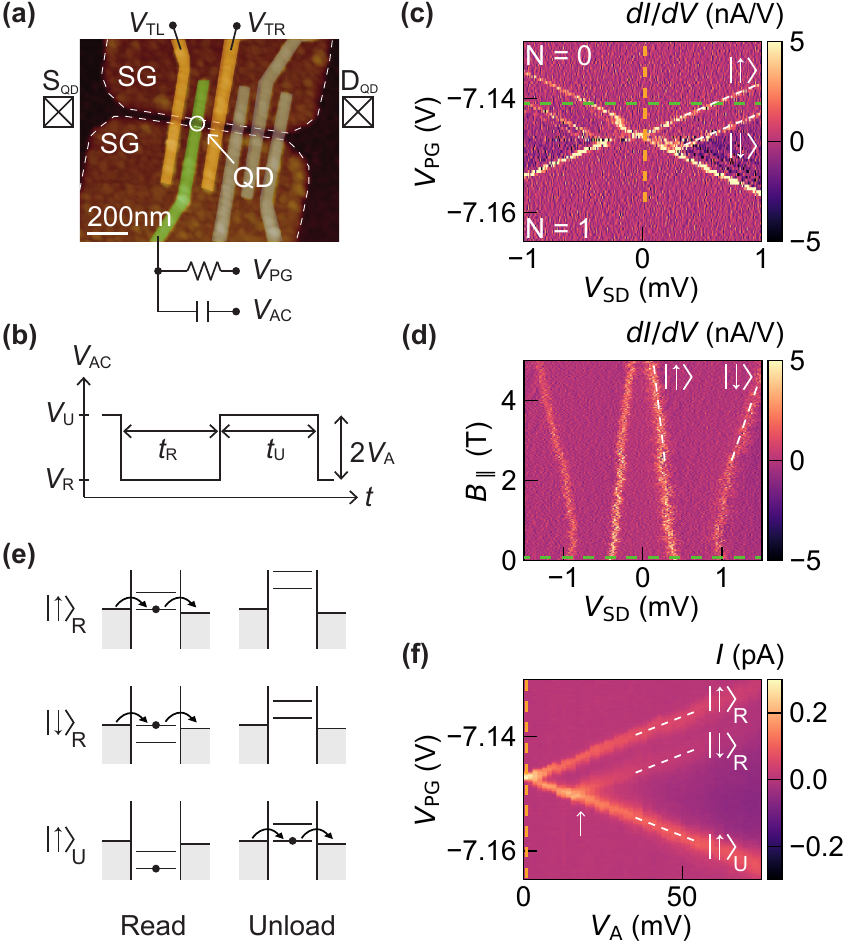}
	\caption{\textbf{(a)} False-color micrograph of the device. The plunger gate (green) is controlled with DC voltage ($V_\mathrm{PG}$) and AC pulses ($V_\mathrm{AC}$) combined by a bias-tee. \textbf{(b)} The pulse sequence is schematically shown with pulse amplitudes $V_\mathrm{R} = -V_\mathrm{A}$ and $V_\mathrm{U} = V_\mathrm{A}$ and pulse widths $t_\mathrm{R}$ and $t_\mathrm{U}$. \textbf{(c)} Coulomb diamond of the first hole transition taken at $B_\mathrm{\perp}=\SI{1.75}{T}$. For large source drain voltage $V_\mathrm{SD}$ the exited state resonance can be observed. \textbf{(d)} In-plane magnetic field dependence at fixed plunger gate voltage, corresponding to green horizontal dashed line in \textbf{(c)}. The splitting of the GS and the ES at high magnetic field is consistent with a spin g-factor of 2. \textbf{(e)} Schematic of transport through $\ket{\uparrow}_\mathrm{R}$ and $\ket{\downarrow}_\mathrm{R}$ during read phase and through $\ket{\uparrow}_\mathrm{U}$ during load phase. For ease of understanding an electron-like level scheme is used, while hole states are probed in the experiment. \textbf{(f)} Coulomb resonance as a function of pulse amplitude at pulse frequency $f=\SI{1.25}{MHz}$. $V_\mathrm{A}=\SI{0}{V}$ corresponds to a line cut along the orange vertical dashed line in \textbf{(c)} at $V_\mathrm{SD}=\SI{40}{\micro eV}$. For finite pulse amplitudes $V_\mathrm{A}$, the transition $\ket{\uparrow}$ splits into two branches, $\ket{\uparrow}_\mathrm{R}$ and $\ket{\uparrow}_\mathrm{U}$, due to the two-level pulsing. For pulse amplitudes $e\alpha_\mathrm{att}\alpha_\mathrm{PG} V_\mathrm{A}>E_\mathrm{Z}$, see white arrow, transport through the excited state is allowed by the pulse during the read phase and shows up as $\ket{\downarrow}_\mathrm{R}$ in between the GS peaks.}
	\label{fig1}
\end{figure}

A false-color atomic force micrograph of the device is shown in Fig.~\ref{fig1}(a). It consists of a hBN encapsulated bilayer graphene flake on top of a global graphite back gate with gold electrodes patterned on top. The gate layers are separated by \SI{30}{nm} atomic-layer-deposited aluminium oxide~\cite{Tong2021}. We form an $n$-type channel connecting source and drain by operating the back gate at $V_\mathrm{BG}=\SI{5}{V}$ and the split gates at $V_\mathrm{SG}=\SI{-3.13}{V}$. Three finger gates are used to control the potential locally along the channel [see Fig.~\ref{fig1}(a)]. The outer two finger gates TL and TR act as tunable tunnel barriers separating the QD from the left and right reservoirs. With increasingly negative voltage applied to the middle finger gate (the plunger gate), we locally lower the Fermi energy set by the back gate, subsequently loading holes into the QD forming below the gate. The sample is mounted in a dilution refrigerator with a base-temperature of~\SI{9}{mK} on a printed circuit board equipped with low-pass-filtered DC lines and $\SI{50}{\Omega}$ impedance-matched AC lines to perform pulsed-gate experiments. The plunger gate is connected to a bias-tee allowing for DC and AC control [e.g., applying a pulse as sketched in Fig.~\ref{fig1}(b)] while all other gates and the ohmic contacts are connected to DC lines only.

We first define our two-level system. Due to spin and valley degrees of freedom in bilayer graphene, the single-particle orbital states are four-fold degenerate~\cite{Garreis2021,Knothe2020}. A perpendicular magnetic field, $B_\mathrm{\perp}$, lifts the degeneracy due to the spin (s) and valley (v) Zeeman effect following $E(B_\mathrm{\perp}) = \pm \frac{1}{2} g_{s,v}\mu_\mathrm{B}B_\mathrm{\perp}$, with the spin $g$-factor $g_s$, the valley $g$-factor $g_v$ and the Bohr magneton $\mu_\mathrm{B}$. Typical values around $g_v = 30$~\cite{Tong2021} ensure that at electron temperatures below \SI{100}{mK}, the two lowest energy levels are valley polarized already at perpendicular magnetic fields $B_\perp > \SI{50}{mT}$. As we operate our device at much higher fields, above $B_\perp = \SI{1.75}{T}$, we can consider our QD as an effective two-level spin system with the ground state (GS) $\ket{K' \uparrow}$ and the excited state (ES) $\ket{K' \downarrow}$ energetically split by $\Delta E= \Delta _\mathrm{SO} + g_s \mu_\mathrm{B}B$ with the zero-field spin orbit splitting $\Delta _\mathrm{SO}$ with typical values on the order of $\SI{60}{\micro eV}$~\cite{Kurzmann2021,Konschuh2012,Huertas-Hernando2006,Banszerus2021Spinvalley}.

To experimentally access the single-particle spectrum we perform finite-bias spectroscopy measurements around the $N=0$ to $N=1$ hole transition at $B_\perp = \SI{1.75}{T}$ [Fig.~\ref{fig1}(c)]. The two lowest $K'$-polarized energy states $\ket{K' \uparrow}$ and $\ket{K' \downarrow}$, from now on denoted as $\ket{\downarrow}$ and $\ket{\uparrow}$ for ease of notation, are well observable while the $K$-states are split off and would only become visible for a larger bias window. 

To confirm the nature of the excited state we apply an additional in-plane magnetic field $B_\mathrm{\parallel}$, which only couples to the spin degree of freedom. Keeping the perpendicular magnetic field $B_\mathrm{\perp}=\SI{1.75}{T}$ and the plunger gate voltage $V_\mathrm{PG}$ fixed [horizontal dashed line in Fig.~\ref{fig1}(c)] we perform finite-bias spectroscopy measurements, varying the in-plane magnetic field $B_\mathrm{\parallel}$ [Fig.~\ref{fig1}(d)]. The splitting $\Delta E$ as a function of magnetic field corresponds to a $g$-factor of 2 and a zero-field spin--orbit splitting $\Delta _\mathrm{SO}\approx \SI{60}{\micro eV}$, consistent with a spin excited state. 

We then perform transient current spectroscopy measurements at a source--drain voltage $V_\mathrm{SD}=\SI{40}{\micro V}\ll k_\mathrm{B}T$ to study the relaxation of the excited state to the ground state. We apply an AC pulse additional to the DC voltage on the plunger gate, effectively shifting the QD states with respect to the Fermi level of the reservoirs in time. We start with two-level pulses each consisting of a read and an unload phase as shown in Fig.~\ref{fig1}(b) with corresponding voltages $V_\mathrm{R}$ and $V_\mathrm{U}$ and pulse widths $t_\mathrm{R}$ and $t_\mathrm{U}$. During $t_\mathrm{U}$, both $\ket{\uparrow}$ and $\ket{\downarrow}$ are pulsed above the Fermi level $E_\mathrm{F}$ of the leads and the QD is emptied. During $t_\mathrm{R}$, if $\ket{\uparrow}$ is aligned with $E_\mathrm{F}$ we observe a steady current. If $\ket{\downarrow}$ is aligned with $E_\mathrm{F}$ during $t_\mathrm{R}$ while $\ket{\uparrow}$ lies below, holes can only tunnel through the QD excited state until one of them relaxes with a spin-flip or until direct tunneling from the leads into the ground state occurs. This effectively blocks transport until the QD gets emptied again in the subsequent unload phase. Fig.~\ref{fig1}(f) shows the current through the QD as a function of pulse amplitude $V_\mathrm{A}$ and $V_\mathrm{PG}$ for $t_\mathrm{U}=t_\mathrm{R}=\SI{400}{ns}$. We observe the splitting of the Coulomb resonance into two peaks, corresponding to transient current through the ground state of the read level, $\ket{\uparrow}_\mathrm{R}$, and the unload level, $\ket{\uparrow}_\mathrm{U}$. At $2V_\mathrm{A} = \SI{100}{mV}$, the peaks are separated by $\Delta V_\mathrm{PG} = \SI{21}{mV}$. This splitting is consistent with the total attenuation of the high-frequency line of our setup. The slope of the ground state splitting, together with the plunger gate lever arm, give us the conversion factor from pulse amplitude to energy scale $\alpha_{\mathrm{att}}\alpha_\mathrm{PG}$, where $\alpha_\mathrm{att}=0.21$ and $\alpha_\mathrm{PG}=0.05$. For pulse amplitudes larger than the Zeeman splitting [indicated in Fig.~\ref{fig1}(f)] the current peak corresponding to a transient current through the excited state of the read level, $\ket{\downarrow}_\mathrm{R}$, becomes visible as well. 
We observe the $\ket{\downarrow}_\mathrm{R}$ peak, because a sufficiently high perpendicular magnetic field $B_\mathrm{\perp}$ is applied. This not only lifts spin and valley degeneracies but also reduces the tunneling rates between the QD and the reservoirs. At low perpendicular magnetic field $\ket{\downarrow}_\mathrm{R}$ is not visible, as the reduction of tunneling rates solely by voltages applied to the tunnel barrier gates TL and TR is not sufficient in this device (see Ref.~\cite{Kurzmann2019Detector}).

\begin{figure}
	\includegraphics{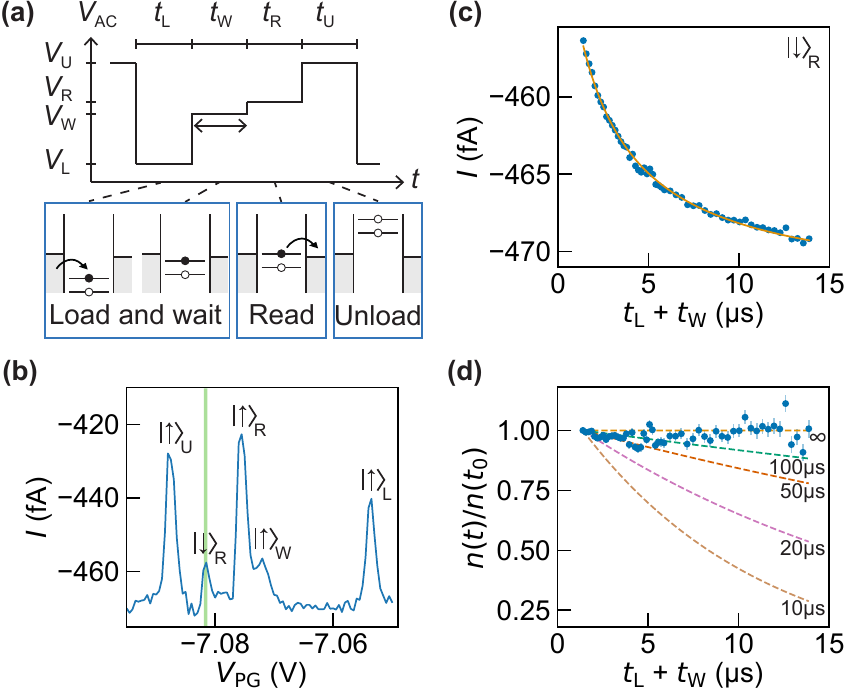}
	\caption{\textbf{(a)} Four-level pulse scheme for relaxation time measurements. The hole enters a random state during the load phase. If the hole is loaded into the GS it will stay there and no transport current can flow. If the carrier is loaded into the ES and does not relax during $t_\mathrm{L}+t_\mathrm{W}$, the charge carrier can tunnel out, resulting in a transport current in the read phase. The QD is then emptied in the unload phase and the cycle is repeated. \textbf{(b)} Current as a function of plunger gate voltage while applying the four-level pulse. We observe various peaks corresponding to GS and ES levels of different pulse phases being aligned with the bias window. This measurement was taken in a second cooldown at a $B_\mathrm{\perp}=\SI{1.9}{T}$, leading to a slight shift of $V_\mathrm{PG}$. \textbf{(c)} Peak corresponding to ES transport during the read phase [shaded in green in \textbf{(b)}] as a function of loading and waiting time. The decay results from a combination of lower cycling rate and excited state relaxation time. As described in main text the decay is dominated by signal decay. \textbf{(d)} Normalized average number of charge carriers per pulse cycle. Dashed color lines represent the decays with the corresponding relaxation time. Our result indicates that relaxation does not play a significant role during the measurement window.}
	\label{fig2} 
\end{figure}

The longer the QD stays in the read phase, the more likely the hole relaxes into the ground state $\ket{\uparrow}$. Therefore, studying the dependence of the amplitude of $\ket{\downarrow}_\mathrm{R}$ on the pulse width $t_\mathrm{R}$ allows us to extract information about the relaxation time. However, two-level pulsed-gate spectroscopy only allows one to extract a lower bound for $T_1$, as this measurement scheme cannot distinguish between relaxation and direct tunneling into the ground state during the read-phase. Already Ref.~\cite{Banszerus2021Pulsed} suffered from this limitation, stating a lower bound of $T_1=\SI{500}{ns}$. To improve on this limitation, we extend the pulsing sequence to four-level pulses with added load and wait phases in which both $\ket{\uparrow}$ and $\ket{\downarrow}$ are pulsed below the bias window~\cite{Fujisawa2002,banszerus2021spin}. The voltage and time for the load phase is chosen such that the time-integral over voltages in one pulse sequence is zero, in order to avoid charging up the bias-tee. The corresponding four-level pulse scheme is depicted in Fig.~\ref{fig2}(a) with pulse widths $t_\mathrm{L}=t_\mathrm{R}=t_\mathrm{U}=\SI{1}{\micro s}$ and $t_\mathrm{W}=\SI{400}{ns}$ and voltages $V_\mathrm{L}=\SI{-9}{mV}$, $V_\mathrm{W}=\SI{0}{V}$, $V_\mathrm{R}=\SI{1.5}{mV}$ and $V_\mathrm{U}=\SI{7.5}{mV}$ for the respective pulse periods. During the load and the wait phases both $\ket{\uparrow}$ and $\ket{\downarrow}$ levels are below the bias window, allowing a hole to tunnel into either one of the two states. During $t_\mathrm{R}$, the excited state $\ket{\downarrow}$ is aligned with the Fermi level of the leads while the ground state $\ket{\uparrow}$ is still below, allowing for spin-selective tunneling. 

Applying the four-level pulsing scheme while sweeping the DC plunger gate voltage $V_\mathrm{PG}$ results in a current trace as shown in Fig.~\ref{fig2}(b). Assigning the observed current peaks to ground state and excited state levels being aligned with the bias window during various pulse phases, we identify the peak $\ket{\downarrow}_\mathrm{R}$ (shaded in green), originating from holes tunneling through the excited state from source to drain during the read phase. If it relaxes to $\ket{\uparrow}$ before the readout, it does not contribute to this current as it cannot tunnel out of the QD, making the amplitude of the current peak a measure of the relaxation. Therefore, we investigate the amplitude of the current peak, $\ket{\downarrow}_\mathrm{R}$, as a function of the waiting time $t_\mathrm{L}+t_\mathrm{W}$ as shown in Fig.~\ref{fig2}(c). We find that the current can be fitted to \cite{Fujisawa2002} $I=\bkt{n} e^{-(t_\mathrm{L}+t_\mathrm{W})/T_1}/t_\mathrm{total} + I_\mathrm{bg}$ in the limit $T_1\rightarrow\infty$, and therefore effectively to $I=\bkt{n}/t_\mathrm{total}+I_\mathrm{bg}$ with high confidence. In these expressions, $\bkt{n}$ is the average number of charge carriers per pulse cycle, $t_\mathrm{total}$ is the total duration of the four-level pulse, and $I_\mathrm{bg}$ is the background current level. This implies that, within our measurement time window the decay of the measured current through $\ket{\downarrow}_\mathrm{R}$ is dominated by the signal strength reducing as the pulse cycle time $t_\mathrm{total}$ becomes longer for increasing $t_\mathrm{W}$. Multiplying $(I-I_\mathrm{bg})$ with the total time we stay below $E_\mathrm{F}$, $t_\mathrm{L}+t_\mathrm{W}$, we find the normalized probability $\bkt{n(t)}/\bkt{n(t_0)}=e^{-(t_\mathrm{L}+t_\mathrm{W})/T_1}$ of the hole still being in the excited state after the load and wait phase as a function of loading and waiting time, $t_\mathrm{L}+t_\mathrm{W}$, shown in Fig.~\ref{fig2}(d). 
Based on this we conservatively estimate a lower bound for the relaxation time $T_1\geq \SI{100}{\micro s}$.
The data was acquired over 8 hours over which the background current fluctuation can be larger than $\SI{0.1}{fA}$ and lead to an uncertainty in the offset current. Careful determination of $I_\mathrm{bg}$ is crucial as it affects the slope in Fig.~\ref{fig2}(d). Further extending the waiting time inevitably leads to vanishingly small current beyond the detection limit of transport measurements. To overcome the signal strength limit, a more advanced readout mechanism is required.

\section{Single-shot excited state detection}

\begin{figure}
    \includegraphics{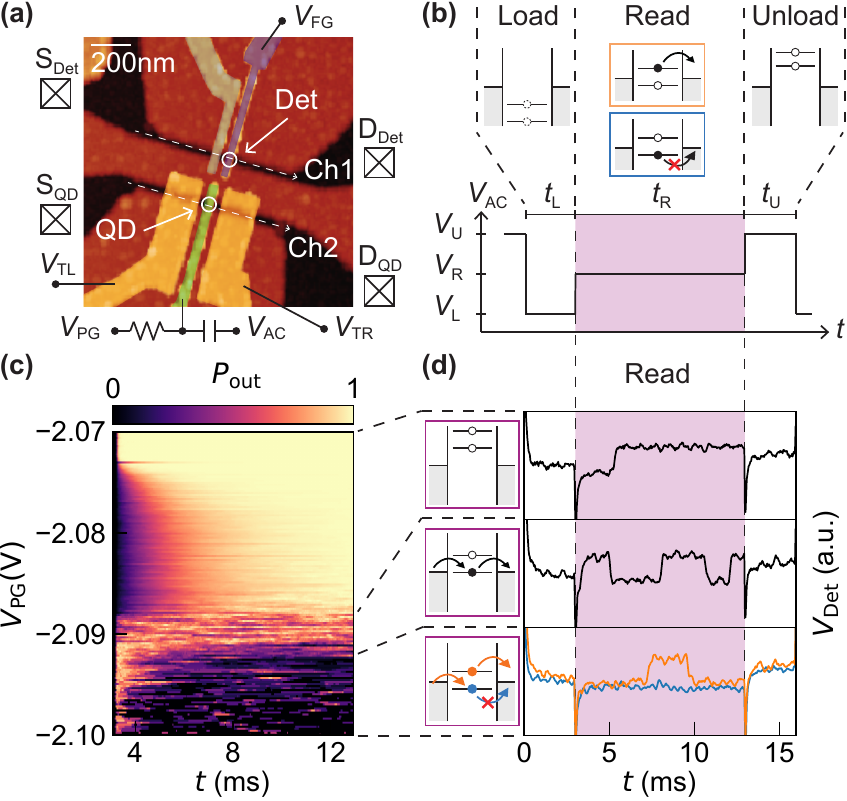}
	\caption{\textbf{(a)} False-color micrograph of a single QD device with a nearby charge detector. The QD is formed beneath the plunger gate (green) in channel 2. TL and TR serve as tunnel barriers to the leads. Another QD is formed under one finger gate (purple) in channel 1 and used as a charge detector. \textbf{(b)} Three-level pulse scheme applied to the plunger gate. \textbf{(c)} Probability of the QD being empty in the read phase as a function of time and plunger gate DC voltage $V_\mathrm{PG}$ extracted from 300 single-shot traces. \textbf{(d)} Exemplary traces for different read levels. Top: Both ES and GS level are pushed above $E_\mathrm{F}$ during the read phase which results in a single tunneling out event. Middle: GS level is aligned with $E_\mathrm{F}$ corresponding to multiple tunneling events. Bottom: Ideal read level configuration allowing for single-shot ES-selective readout.}
	\label{fig3} 
\end{figure}

Recent progress in fabrication techniques enabled the realization of a fully electrostatically defined device with an integrated charge detector as described in Ref.~\cite{Kurzmann2019Detector}. The device shown in Fig.~\ref{fig3}(a) consists of two channels separated by a depletion region underneath a \SI{150}{nm} wide middle gate. The QD is formed beneath the plunger gate in channel 2 whereas TL and TR serve as tunnel barriers to the leads as in the previous sample. The charge detector is based on a second QD formed below a single finger gate (FG) in the current biased channel 1. A charge carrier tunneling on or off the QD in channel 2 changes the electrostatic potential in its surroundings, therefore also in the nearby sensing QD in channel 1. This potential change shifts the Coulomb resonances in the sensing QD and leads to a step-like change in the voltage across channel 1, if the sensing QD is tuned to the steep slope of a conductance resonance.

At a perpendicular magnetic field of \SI{2}{T}, we find an excited state split from the ground state by about $\SI{300}{\micro eV}$. 
We exclude an orbital excitation, as the expected orbital splitting is a few $\SI{}{meV}$. 
With an out-of-plane magnetic field of $\SI{2}{T}$, typical valley splittings are $\SI{0.6}{meV}$ - $\SI{6}{meV}$, with a valley $g$-factor of 10 - 100~\cite{Tong2021}. 
Therefore, the excited state is probably a spin state with $g\mu_\mathrm{B} B=\SI{232}{\micro eV}$ and a spin-orbit coupling $\Delta_\mathrm{SO} \approx \SI{68}{\micro eV}$~\cite{refnote_state2021}. 

For the single-shot charge detection experiment, we follow the scheme introduced by Elzerman in Ref.~\cite{Elzerman2004}. Fig.~\ref{fig3}(b) shows the voltage pulses applied to the plunger gate PG.
Again, in the load phase either the GS or the ES can be loaded as both energy levels reside below the Fermi energy $E_\mathrm{F}$ of the leads. During the loading time $t_\mathrm{L}$ the charge carrier is trapped on the QD and Coulomb blockade prevents an additional hole from entering. After $t_\mathrm{L}$ the pulse amplitude is changed to $V_\mathrm{R}$ such that the energy level of the ES is pushed above $E_\mathrm{F}$ while the GS level remains below. Therefore, the charge carrier can only tunnel off the QD if it was loaded onto the ES in the previous phase. Once the charge carrier has left the QD from the excited state, the Coulomb blockade is lifted and another charge carrier can tunnel into the GS. The combination of these two processes, tunneling out from the excited state, followed by tunneling into the ground state, leads to a characteristic `blip' in the detector voltage, indicative of excited state loading during the load phase. Note that here we pulse the ES above $E_\mathrm{F}$ of the leads, in contrast to the transport measurements before, where we aligned the ES level with the leads during the read phase. After $t_\mathrm{R}$ we enter the unload phase, where the pulse amplitude is changed to $V_\mathrm{U}$, and both energy levels are pushed above $E_\mathrm{F}$ emptying the QD. As a response to the three-level pulse we expect a change of voltage across the sensing QD consisting of two contributions. First, due to capacitive coupling between PG and the sensing QD the voltage will change proportionally to the applied pulse amplitude. Second, the voltage across the sensing QD traces the charge occupation of the QD, stepping up or down as soon as a charge carrier tunnels on or off the QD. Whenever we load the GS during load phase, the voltage trace should stay flat during $t_\mathrm{R}$. Therefore, measuring whether the voltage trace shows a blip or not during the read out phase forms the basis of our single-shot readout mechanism.

\begin{figure}
    \includegraphics{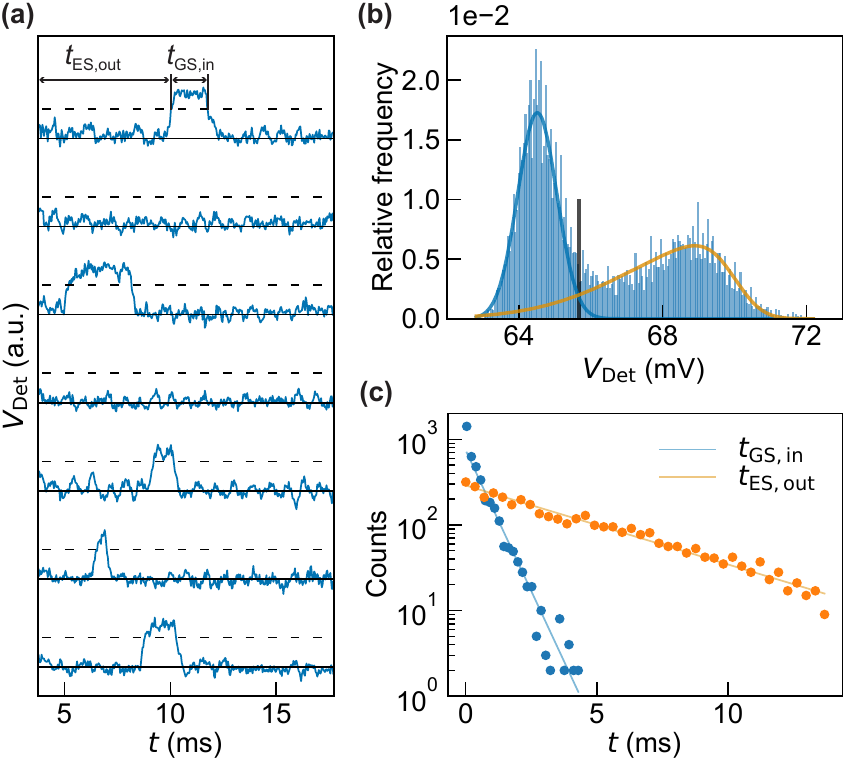}
	\caption{Readout statistics. \textbf{(a)} Exemplary single-shot readout traces each shifted by $\SI{12}{mV}$ for clarity. If the voltage passes the threshold value (dashed lines) it indicates an excited state was loaded in the previous load phase. \textbf{(b)} Histogram of peak values in the read phase. The two peaks are fitted with skewed Gaussians, represented in blue and orange. The visibility $V=87.1\%$ is obtained by setting an optimal threshold as marked by the black line. \textbf{(c)} Histogram of tunnel-in and -out times as defined in \textbf{(a)}. The transparent lines are exponential fits to the histogram. We extract an ES tunnel-out rate of $\SI{4.64}{ms}$ and a GS tunnel-in rate of $\SI{0.66}{ms}$.}
	\label{fig4} 
\end{figure}

To find the appropriate alignment of the read-level for ES-dependent tunneling we sweep $V_\mathrm{PG}$ while keeping the three-level pulse amplitude constant. We choose pulse widths $t_\mathrm{L}=t_\mathrm{U}=\SI{3}{ms}$ and $t_\mathrm{R}=\SI{10}{ms}$, long enough to allow the charge carrier to tunnel in and out of the QD. The pulse voltages $V_\mathrm{L}$ and $V_\mathrm{U}$ ensure that both energy levels are pushed below (above) $E_\mathrm{F}$ of the leads during the load (unload) phase. Starting with $V_\mathrm{PG}$ being too high, both GS and ES levels lie above $E_\mathrm{F}$, such that the charge carrier can always tunnel off the QD regardless of being in the ES or the GS. The characteristic voltage trace in this regime only shows a step up corresponding to the unloading event, and then remains at this higher level as shown in Fig.~\ref{fig3}(d) top panel. When lowering $V_\mathrm{PG}$ we enter the regime of random telegraph signal corresponding to the GS being aligned with $E_\mathrm{F}$ of the leads such that a charge carrier can tunnel on and off the QD multiple times within the read phase as observed in Fig.~\ref{fig3}(d) middle panel. Further decreasing $V_\mathrm{PG}$ we find the correct read level where we can distinguish between the GS and ES of the QD being occupied. A single blip at the beginning of the read phase corresponds to the ES tunneling off the QD and subsequent tunneling into the GS from the leads. A flat trace during $t_\mathrm{R}$ indicates that the charge carrier was loaded into the GS during the load phase, and is thus trapped on the QD. Alternatively, the excited state could have been loaded, but it relaxed into the ground state, before it could tunnel out. In this region the condition $\mu_\mathrm{ES} > E_\mathrm{F} > \mu_\mathrm{GS}$ is fulfilled and we can perform a single-shot projective measurement of the state of the charge carrier. Exemplary traces for the two cases are shown in Fig.~\ref{fig3}(d) bottom panel, with the blue trace corresponding to the GS and the orange one with the characteristic 'blip' corresponding to the ES.

Averaging the read-phase voltage across the sensing QD $\bkt{V_\mathrm{Det}}$ over 300 single-shot traces at different plunger gate voltages $V_\mathrm{PG}$ we estimate the probability of the QD being empty as a function of time for each cycle as shown in Fig.~\ref{fig3}(c). The correct readout configuration shows a higher current in the beginning of the read phase, as observed in the lower part of Fig.~\ref{fig3}(c). The excited state tunneling events can only be observed if the excited state does not decay into the ground state during the time it stays in the load phase $t_\mathrm{L}=\SI{3}{ms}$. Thus, observing a 'blip' makes us conclude that the relaxation time is on the order of milliseconds.

Typical experimental traces of the detector response in the read phase are shown in Fig.~\ref{fig4}(a). If the trace surpasses the threshold voltage (dashed line) within the read phase window, it indicates that the hole occupied the ES during the previous load phase. Analyzing a data set consisting of 10'000 single-shots we extract the histogram of the peak values of $V_\mathrm{Det}$ shown in Fig~\ref{fig4}(b). We identify two well-defined peaks indicating that the detector voltage $V_\mathrm{Det}$ has two favorable values corresponding to an additional hole being in the QD or not. By further analyzing the traces with a 'blip', we extract the tunnel-out time of the ES $\tau_\mathrm{ES,out}=\SI{4.64}{ms}$ and the tunnel-in time of the GS $\tau_\mathrm{GS,in}=\SI{0.66}{ms}$ with fits to the data shown in Fig~\ref{fig4}(c). 

By choosing a threshold that optimizes the readout visibility~\cite{Keith_2019,Morello2010,Elzerman2004}, we obtain the ground state readout fidelity $F_\mathrm{GS} = 97.0\%$, the excited state readout fidelity $F_\mathrm{ES} = 90.1\%$, and the visibility $V=F_\mathrm{ES}+F_\mathrm{GS}-1 = 87.1\%$. The readout fidelity is limited by finite measurement bandwidth and charge sensitivity. An improved device geometry could include a wider detector channel for higher detector conductance and an optimized ratio between channel separation and hBN thickness for stronger capacitive coupling between the charge sensor and the QD. Faster charge detection can be achieved by using a rf-charge detector~\cite{Reilly2007,Noiri2020,Barthel2010}, dispersive charge sensing by coupling the QDs to lumped-element LC tank circuit~\cite{Betz2015,Colless2013,House2015,Crippa2019}, or by an on-chip superconducting resonator~\cite{Samkharadzeeaar4054,Zheng2019,Mi2018}.

\section{Conclusion}

In conclusion, we first characterized the lifetime of excited states in a graphene QD through direct transport measurements. The measurement is limited by signal strength rather than the relaxation time, comparable to previous measurements performed in bilayer graphene QDs. 
In a second step we presented a QD device integrated with a capacitively coupled charge sensor capable of resolving a single charge tunneling event in the time-domain. The sensitivity of the charge detector enables excited state readout in the single-shot limit - an essential step towards a fully controllable quantum processor in graphene. Our result further indicates that the relaxation time is of the order of milliseconds, comparable to most semiconductor qubits encoded by electron spin, showing that spin qubits in bilayer graphene QDs are a promising platform.

\section{Acknowledgement}
We are grateful for the technical support from Peter M\"arki, Thomas B\"ahler and the staff of the ETH FIRST cleanroom facility. We thank Luca Banszerus and Christian Volk for discussions. We acknowledge financial support by the European Graphene Flagship, the ERC Synergy Grant Quantropy, the European Union’s Horizon 2020 research and innovation programme under grant agreement number 862660/QUANTUM E LEAPS and NCCR QSIT (Swiss National Science Foundation) and under the Marie SKlodowska-Curie Grant Agreement Number 766025. KW and TT acknowledge support from the Element Strategy Initiative conducted by the MEXT, Japan, Grant Number JPMXP0112101001, JSPS KAKENHI Grant Number JP20H00354 and the CREST(JPMJCR15F3), JST. 


\begin{thebibliography}{47}
\bibitem [{ref()}]{refnote_state2021}%
  \BibitemOpen
  \href@noop {} {}\bibinfo {note} {For a detailed analysis of the nature of the
  excited state, magnetic field-dependent finite-bias measurements are
  required. They are challenging, however, due to the low tunneling rate
  through the QD and its non-monotonic dependence on magnetic field. Retuning
  the device will be necessary at each magnetic field to make the excited state
  visible, as magnetic fields shift both the Coulomb resonance of the charge
  sensor and the tunneling rate.}\BibitemShut {Stop}%
\end{thebibliography}

%

\end{document}